# UNLOCKING THE FUTURE OF FOOD SECURITY THROUGH ACCESS TO FINANCE FOR SUSTAINABLE AGRIBUSINESS PERFORMANCE

**Ayobami Paul Abolade \***
*Department of Business & Entrepreneurship*
*Kwara State University*
*Malete, Nigeria*
*ORCID 0000-0001-9302-7189*

**Ibrahim Olanrewaju Lawal**
*Department of Business & Entrepreneurship*
*Kwara State University*
*Malete, Nigeria*

**Kamoru Lanre Akanbi**
*Department of Business & Entrepreneurship*
*Kwara State University*
*Malete, Nigeria*

**Ahmed Orilonise Salami**
*Africa Business School,*
*UM6P,*
*Rabbat, Morroco*

*\*Corresponding author email:* ayoabolade@gmail.com

**Abstract.** Access to finance is vital for improving food security, particularly in developing nations where agricultural production is crucial. Despite several financial interventions targeted at increasing agricultural production, smallholder farmers continue to lack access to reasonable, timely, and sufficient financing, limiting their ability to invest in improved technology and inputs, lowering productivity and food supply. This study examines the relationship between access to finance and food security among smallholder farmers in Ogun State, employing institutional theory as a theoretical framework. The study takes a quantitative method, with a survey for the research design and a population of 37,200 agricultural smallholder farmers. A sample size of 380 was chosen using probability sampling and simple random techniques. The data were analysed via Partial Least Squares Structural Equation Modelling (PLS-SEM). The findings demonstrate a favourable relationship between access to finance and food security, with an $R^2$-value of 0.615 indicating a robust link. These findings underline the need of improving financial institutions and implementing enabling policies to enable farmers have access to the financial resources they need to achieve food security outcomes.

## INTRODUCTION

Agribusiness is crucial to promoting food security, economic growth, and sustainable development, especially in developing nations. The agricultural sector not only contributes significantly to the GDP, but it also employs a large number of the people, particularly in rural areas. In terms of food security, agribusiness contributes to the availability, accessibility, and affordability of food, which is critical for the survival and well-being of the populace. However, despite agribusiness's critical role in providing food security, numerous hurdles limit the sector's ability to fulfil rising food demand. One of the most significant issues is limited access to finance. Smallholder farmers, who make up a sizable share of agribusiness entrepreneurs, frequently struggle to secure the capital needed to invest in productive technology, buy inputs, and scale their operations. Without enough capital, their ability to increase output, implement modern farming techniques, and develop their enterprises is severely limited.

Access to finance is critical for increasing the performance of agribusinesses and maintaining food security. Adequate finance allows farmers and agribusinesses to invest in improved seeds, fertilisers, irrigation systems, mechanisation, and post-harvest technology. It also enables them to withstand economic downturns, adjust to climate change, and capitalise on new market opportunities. In contrast, agribusinesses' ability to grow is hampered by a lack of access to financing and financial services, resulting in negative repercussions on food production, availability, accessibility, affordability, and distribution.

Smallholder farmers frequently fail to receive loans from commercial banks due to a number of factors, including high interest rates, collateral requirements, and a lack of financial literacy. Furthermore, financial institutions view agriculture as a high-risk sector due to its susceptibility to climate change, pests, and market volatility. These issues are worsened by poor institutional frameworks, insufficient government support, and a lack of appropriate finance products for the agriculture industry.

The availability of funds has a direct impact on food security. Agribusinesses can use financial resources to boost productivity, eliminate post-harvest losses, and improve the efficiency of food supply networks. This, in turn, serves to guarantee that food is available in sufficient quantities, accessible to all parts of the population, and reasonably priced for customers. In many developing nations, particularly in Sub-Saharan Africa, food insecurity is a chronic problem. Limited access to finance exacerbates the situation, as smallholder farmers and rural entrepreneurs are frequently unable to invest in technology that would increase yields and stabilise the food supply. Furthermore, inability to scale operations frequently results in lower production outputs and inefficient distribution systems, limiting food availability in both local and national markets. However, the post-COVID-19 epidemic has uncovered further vulnerabilities in the farming industry. Disruptions in supply networks, employment shortages, and reduced earnings have exacerbated food insecurity. In this setting, enhancing agriculture access to finance is even more crucial for promoting recovery, increasing productivity, and strengthening food systems. This study will look at how access to finance affects agribusiness performance.

**LITERATURE REVIEW**

1. **Access to Finance**

Access to finance refers to the ease with which individuals or businesses can obtain formal financial products and services, such as credit, savings, insurance, and payment systems, required for economic participation, as well as the ability of businesses or individuals to secure necessary funding or credit at reasonable interest rates without excessive bureaucratic hurdles or discriminatory practices (Demirgüç-Kunt et al., 2020; Asongu & Nwachukwu, 2021).

Furthermore, access to finance describes a financial system that inclusively accommodates underserved or vulnerable populations, such as low-income households, rural enterprises, and women, ensuring that financial services are available to all, and it involves financial institutions' ability to facilitate the provision of loans, credit facilities, and investment capital to businesses, particularly small and medium-sized enterprises (SMEs), with minimal risk-related (Kabeer et al., 2021; Beck & Maimbo, 2021). Furthermore, access to finance entails the utilisation of digital financial platforms, such as mobile banking and digital wallets, allowing individuals and businesses to participate in financial systems without the need for physical banking infrastructure, and it refers to the availability of financial services that are specifically designed to meet the unique needs of certain sectors or groups, such as microfinance for small-scale farmers or SMEs, which frequently face barriers in traditional lending models (Suri & Jack, 2021; Grohmann, 2021).

Furthermore, access to finance entails ensuring that financial resources are provided on sustainable terms, which means that loans and other forms of credit are not only available but structured in a way that borrowers can repay them without causing financial distress, and it is characterised by the reduction of common barriers such as high collateral requirements, high interest rates, or complex application processes, allowing easier entry for businesses and individuals. (Akinwumi et al., 2023; Omondi & Mugambi, 2023). Access to finance is the process by which financial institutions provide adequate risk mitigation mechanisms, such as credit guarantees or insurance products, to reduce the perceived risks of lending to certain sectors, particularly agriculture or SMEs. It also includes the enhancement of financial literacy, which empowers individuals and businesses to make informed decisions about credit, savings, and investment opportunities, increasing their ability to participate in financial markets (Asogwa & Nwaogwugwu, 2022; Chen et al., 2022).

Access to finance is broadly described as individuals and organisations' capacity to receive relevant financial products and services from formal financial institutions, such as credit, savings, insurance, and payment systems (Demirgüç-Kunt et al., 2020). Access to these resources is critical for business growth, risk management, and investment in new opportunities. The World Bank (2020) emphasises that without proper financial access, firms, particularly SMEs and agricultural sectors, are frequently unable to develop or enhance efficiency, contributing to economic stagnation.

### 1.1 Determinants of Access to Finance

Several factors influence financial access, including institutional frameworks, financial literacy, digital infrastructure, and financial institutions' risk perceptions. Beck and Maimbo (2021) identify the regulatory environment as a crucial predictor, pointing out that countries with stronger legislative frameworks for financial inclusion have higher levels of loan availability. Similarly, Asongu and Nwachukwu (2021) contend that corruption, bureaucratic inefficiency, and weak property rights severely hinder the availability of funding to SMEs in many emerging nations. Financial literacy is another important driver, as it influences whether individuals and businesses can obtain loans and financial services. Grohmann (2021) discovered that better levels of financial literacy among SMEs enhanced their ability to acquire formal loans and financial services, therefore supporting business growth.

#### 1.1.1 Challenges of Access to Finance

Despite global attempts to enhance financial inclusion, major hurdles still exist, particularly for marginalised communities and rural businesses. According to Asogwa and Nwaogwugwu (2022), main barriers to rural company financing include high interest rates, onerous collateral requirements, and a lack of specialised financial solutions. Furthermore, commercial banks view agribusiness and SMEs as high-risk due to their vulnerability to market volatility and environmental concerns, making financial institutions hesitant to lend to them (Omondi & Mugambi, 2023). Furthermore, gender discrepancies in financial access continue to exist. According to Kabeer et al. (2021), cultural norms, a lack of collateral, and low financial knowledge make it more difficult for women entrepreneurs to secure loans, particularly in Sub-Saharan Africa and South Asia. These inequities prevent women-led firms from growing, limiting their economic impact.

#### 1.1.2 Digital Finance and Access to Finance

The rise of digital banking has created new opportunities to improve financial inclusion. Mobile banking, digital wallets, and Fin-tech advances are revolutionising access to finance, especially in areas with limited penetration of traditional banking institutions. According to Suri and Jack (2021), mobile money platforms have played an important role in improving financial access in Africa by allowing individuals and businesses to save, borrow, and transact without the need for traditional banks.

Akinwumi et al. (2023) found that digital finance has lowered transaction costs and expanded access to microfinance for SMEs in East Africa. Furthermore, digital lending platforms have made it simpler for small enterprises and farmers to obtain short-term loans without providing physical collateral (Wamba et al., 2022). However, issues such as digital literacy, cybersecurity concerns, and regulatory deficiencies persist (Chen et al., 2022).

### 1.2 Agribusiness Performance

Agribusiness, defined as the interconnected sectors of agriculture, food production, and related services, is critical to global economic development, food security, and rural lifestyles. This literature study focusses on conceptual definitions of agricultural performance, with a particular emphasis on food security and other aspects of performance. Agribusiness performance is a multidimensional notion that incorporates many aspects of efficiency, effectiveness, and sustainability in the agricultural industry.

Kueh et al. (2021) define agribusiness performance as the ability of firms within the agricultural value chain to fulfil their strategic goals while maximising resource utilisation. This encompasses not only financial indicators like profit margins and ROI, but also operational efficiency, customer happiness, and environmental stewardship. Several researchers emphasise the relevance of integrating several performance variables. For example, Kariuki et al. (2022) recommend that agribusiness performance be evaluated using a balanced scorecard approach that takes into account financial performance, internal procedures, customer viewpoints, and learning and growth. This complete view enables stakeholders to evaluate how well agribusinesses respond to changing market conditions while meeting long-term sustainability objectives.

#### 1.2.1 Food Security

Food security is the state in which all individuals have consistent physical and economic access to sufficient, safe, and nutritious food that meets their dietary requirements for a healthy and active life (FAO, 2020). Food security is described by four pillars: availability, access, utilisation, and long-term stability. These pillars work together to ensure that individuals and households can meet their food needs consistently and without disruption (Clapp, 2021). Global food security refers to the ability of worldwide and national food supply systems to withstand and recover from economic, environmental, and health crises, providing continued food access (Laborde et al., 2020). Food security at the home level refers to all members of a household having consistent access to the quantity and quality of food needed to live an active and healthy life (Renzaho, 2020). Food security is the state in which people have access not only to enough calories but also to nutritious food, which promotes good health and prevents malnutrition or diet-related disorders (Swinburn et al., 2020).

Food security includes both economic and social factors, such as price and cultural tolerance of food. It guarantees that food systems deliver adequate quantities of culturally acceptable foods that are consistent with population tastes and social norms (Béné et al., 2021). Food security entails the long-term viability of food systems, which means that current food production and consumption patterns must not jeopardise future generations' access to sufficient and healthy food (Ingram, 2021).

Food security is inextricably related to environmental stability, as climate change, soil degradation, and water shortages all have an impact on food production and distribution, influencing long-term food availability and access. Food security is defined as a state in which policy measures ensure that food systems are resilient to shocks such as pandemics or conflicts that may disrupt food availability or access on a national or global scale (Chauvin et al., 2022). Food security is increasingly being viewed as a human right, with individuals having an intrinsic right to adequate food and governments being required to develop policies that ensure this right for all citizens (De Schutter & Giljum, 2021).

Food security, a multidimensional term, is constantly evolving in academic and policy discourse. The Food and Agriculture Organisation (FAO) defines a state as one in which "all people, at all times, have physical, social, and economic access to sufficient, safe, and nutritious food that meets their dietary needs and food preferences for an active and healthy life" (FAO, 2020). Recent research (2020-2023) has focused on the key characteristics of food security, including availability, access, utilisation, and stability. Food availability is the actual presence of food in a certain location, encompassing production, importation, and distribution (Clapp, 2021).

Climate change, population increase, and conflict have had a substantial impact on the global food supply, resulting in variations in agricultural output (Wheeler & von Braun, 2022). As agricultural systems cope with the negative effects of climate change, such as droughts and floods, global food production faces substantial problems, affecting food availability in vulnerable areas.

In Sub-Saharan Africa, a report from the Famine Early Warning Systems Network (FEWS NET, 2022) found that climate-related shocks worsened food scarcity, affecting production and supply systems and endangering food availability. Access to food includes both economic and physical access, ensuring that people can get the food they need (Renzaho, 2020). Economic issues such as income, inflation, and food costs have a direct impact on households' ability to obtain adequate food. For example, COVID-19 exacerbated global food insecurity by disrupting household incomes and market access (Béné et al., 2021).

Millions of people have more difficulties acquiring food in areas where food prices soared, such as the Middle East and Sub-Saharan Africa (Béné, 2020). Furthermore, physical access has been limited due to infrastructure concerns and conflict zones, notably in war-torn countries like Yemen and Ukraine (Global Network Against Food Crises, 2022). The idea of utilisation in food security refers to the nutritional and food safety components, which ensure that food consumption leads to beneficial health outcomes (Ingram, 2021). This dimension encompasses dietary diversity, nutritional value, and the hygiene circumstances in which food is produced and consumed. The double burden of malnutrition, characterised by both undernutrition and overnutrition, has emerged as a significant concern in current talks about food utilisation.

According to Swinburn et al. (2020), global food systems have evolved towards the use of highly processed, energy-dense foods, which has contributed to an increase in diet-related noncommunicable diseases such as diabetes and obesity, as well as persistent malnutrition in many low-income nations. Addressing the nutritional quality of food has become crucial in ensuring that people not only have access to food, but also that the food they eat promotes overall health. Stability refers to food systems' ability to provide consistent access to food over time, despite of external shocks such as economic crises, pandemics, or natural disasters (FAO, 2021). This component of food security has received more attention in the aftermath of global disruptions such as COVID-19, which exposed flaws in food supply chains (Laborde et al., 2020).

The epidemic impacted worldwide logistics and agricultural output, resulting in price increases and supply shortages. Furthermore, the Russia-Ukraine war, a major component in global wheat supply, has disrupted grain exports, threatening food security in import-dependent countries such as Africa and the Middle East (Chauvin et al., 2022). Stability also includes the resilience of food systems to future shocks, which necessitates adaptive solutions to protect food production and delivery.

### 1.2.2 Operational Efficiency

Operational efficiency is another important aspect of agricultural performance. It refers to the efficiency with which resources are used to produce goods and services. Balogun et al. (2021) define operational efficiency as agribusinesses' ability to optimise production processes, reduce waste, and improve supply chain management. Efficient operations not only decrease costs, but also improve product quality, resulting in increased customer satisfaction. Recent research emphasises the importance of technology and innovation in improving operational efficiency in agribusiness.

For example, Uche et al. (2023) state that the use of digital technologies, precision agriculture, and data analytics can considerably boost productivity and operational efficiency. Agribusinesses can use technology to streamline processes, manage resources more effectively, and respond quickly to market demands.

### 1.2.3 Sustainability and Environmental Impact

Sustainability has evolved as an important part of agribusiness success, reflecting the sector's broader commitment to environmental conservation and social equality. According to Fadahunsi and Adegboye (2023), sustainable agriculture practices include using natural resources responsibly, reducing environmental degradation, and improving social welfare. This aspect of performance is becoming increasingly relevant as consumers and stakeholders demand greater accountability for the environmental impact of agricultural practices. Furthermore, incorporating sustainability criteria into agricultural performance evaluations can increase long-term viability.

According to a study by Nwankwo and Udo (2022), organisations that embrace sustainable practices generally profit from enhanced brand recognition, customer loyalty, and access to new markets. As such, sustainability is more than just an ethical consideration; it is a strategic need for improving agribusiness success.

**2 Access to Finance and Agribusiness Performance: An Institutional Theory Perspective**

Access to financing is a significant factor in agribusiness performance, influencing investment decisions, operational efficiency, and overall sustainability. The concept of institutional theory, as stated by North (1990), provides a strong lens for understanding how agribusinesses handle financial limitations and improve performance. This paper examines the relationship between access to financing and agribusiness success through the lens of institutional theory, emphasising both traditional and innovative institutional viewpoints.

**i. Old Institutionalism and Agribusiness Financing**

Old institutionalism, which has its roots in the works of Stinchcombe (1997) and Scott (1987), emphasises how organisations can adapt to environmental conditions while maintaining institutional ideals. In the context of agriculture, access to finance is a critical resource that is influenced by institutional arrangements such as government policies, banking regulations, and market structure. Agribusinesses frequently respond to these constraints by establishing tactics that are consistent with the prevalent norms and values in their institutional environment. For example, resource dependence theory, a core component of traditional institutionalism, holds that organisations must obtain crucial resources, such as financial capital, in order to exist and grow (Pfeffer and Salancik, 1978).

Agribusinesses operating under this framework are required to establish partnerships with financial institutions and align their operations with their expectations and standards. This improves their access to capital, which is critical for investing in manufacturing, technology, and infrastructure. Furthermore, contingency theory, which examines the impact of external circumstances on organisational performance, emphasises the significance of aligning financial strategy with environmental conditions (Lawrence & Lorsch, 1967).

Agribusinesses must tailor their financial processes to the unique problems they confront, such as fluctuating commodity prices and shifting consumer tastes. This adaptation illustrates the ongoing evolution of organised organisational actions in response to financial constraints while adhering to institutional principles.

**ii. New Institutionalism and Legitimacy in Financial Access**

In contrast, modern institutionalism emphasises how organisations gain resources by adhering to normative norms and developing reputation in their domains (Greenwood & Hinings, 1996). The perception of legitimacy among financial institutions, investors, and stakeholders has a significant impact on agriculture financing.

Organisations that demonstrate conformity with regulatory requirements, industry norms, and best practices are more likely to receive financial assistance. Meyer and Rowan (1997) contend that organisations frequently increase their legitimacy by matching their activities with collectively valued goals. This alignment can take many forms, including adherence to sustainable agriculture methods, community engagement, and social responsibility activities. Agribusinesses can improve their reputation and attract possible investors by demonstrating their dedication to these ideals.

DiMaggio and Powell (1983) also establish the idea of isomorphism, which describes how organisations in the same field become more similar over time as they respond to normative forces.

Agribusinesses that meet the expectations of financial institutions may have an easier time obtaining capital, as lenders prefer organisations that follow established operating procedures. This is especially important in situations where financial institutions prioritise investments in agribusinesses that match with sustainability objectives or demonstrate social effect.

### iii. The Integration of Old and New Institutionalism

Oliver (1997) recognises the importance of combining both old and contemporary institutional approaches, emphasising that legitimacy gained via social acceptance, along with economic optimisation of structures and processes, adds to organisational success. For agribusinesses, integration entails not just responding to financial restrictions but also establishing legitimacy in the eyes of stakeholders. Agribusinesses can improve their performance and sustainability by combining financial optimisation with a commitment to social and environmental values. In this context, institutional theory investigates the framework and extent to which agribusinesses respond to financial rules and regulations.

This examination allows for a better understanding of how financial access influences the Overall performance of agribusinesses in the larger ecosystem. Agribusinesses who successfully traverse this landscape will be better positioned to acquire capital, generate innovation, and respond to market demands, thereby increasing their competitive edge.

### 3. Empirical Review

According to Adebayo et al. (2021), improved access to financing considerably enhances smallholder farmers' ability to invest in better agricultural methods. The study found that farmers who used financial services had higher crop yields and better food availability, showing a direct link between financial access and food security.

Asfaw et al. (2020) argue that access to financial services enables farmers to purchase required inputs such as seeds and fertilisers. Their findings showed that enhanced liquidity allows farmers to manage the risks associated with agricultural production, resulting in better food security. This access is essential for farmers to respond effectively to market swings and production constraints. Sulaiman et al. (2022) claimed that financial inclusion measures greatly reduce food insecurity. Access to finance improves food security by allowing households to engage in productive activities and diversify their sources of income. The findings highlighted the importance of financial services in developing resistance to food insecurity.

Furthermore, Zougmoré et al. (2021) did a study in West Africa and discovered that farmers who had access to microfinance institutions were better prepared to invest in climate-friendly agricultural techniques. This financial assistance enabled them to boost agricultural output and improve food security in their communities, illustrating the importance of access to finance in promoting sustainable agricultural practices. Furthermore, Nugroho et al. (2022) investigated the effect of financial literacy on food security in Indonesia. Their findings demonstrated that farmers with better levels of financial literacy were more likely to use financial products and services, which improved their food security situation. This emphasises the role of financial knowledge and access to financing in increasing food security. Kumar and Singh (2023) also look at the role of agricultural cooperatives in enhancing farmers' access to financing. The study discovered that cooperative membership considerably increased farmers' access to finance and insurance goods. As a result, this access boosted their food production capacity and helped to increase food security in rural areas.

### METHODOLOGY

This study investigated the impact of access to finance on agricultural company performance in Ogun State. 380 respondents were chosen from a population of 37,200 rice farmers through probability sampling, using simple random selection procedures. Primary data was used as a source of data; a questionnaire was constructed to collect data. However, quantitative analysis was performed to assess the extent to which explanatory variables (access to finance) influence the provided variable (food security). Pls-Sem was utilised for data analysis, and regression and path measurement were investigated.

**Data Analysis and discussion of Findings**

**Hypothesis testing: Access to finance and Food Security**

**H$_{O1}$:** there is no significant relationship between access to finance and food security

This predicts the relationship between the explanatory variable (*Access to finance*) and the given variable (*Food security*). The result of hypothesis is as follows:

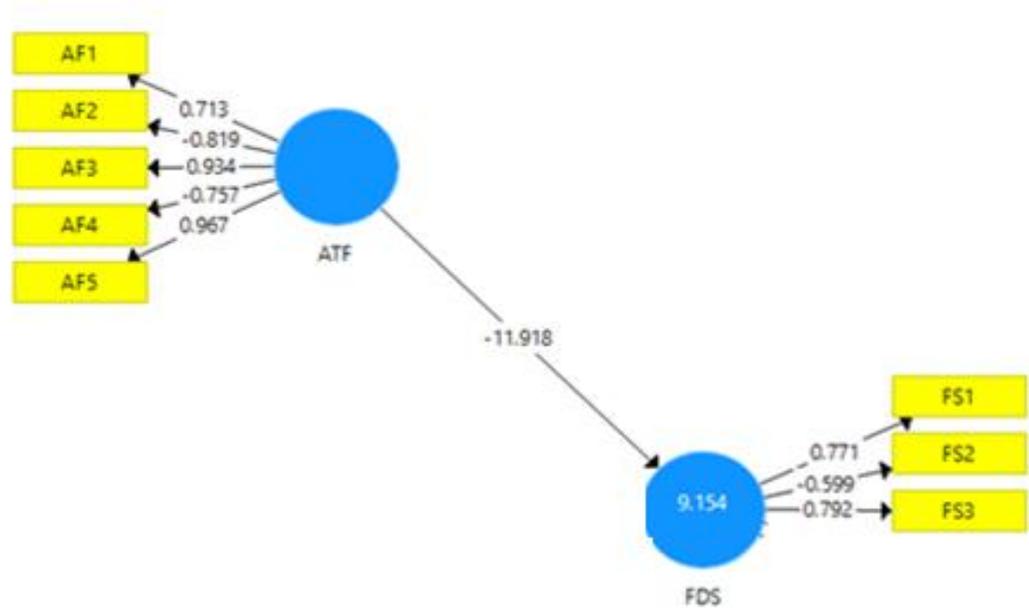

*Fig. 1 Measurement of variables*

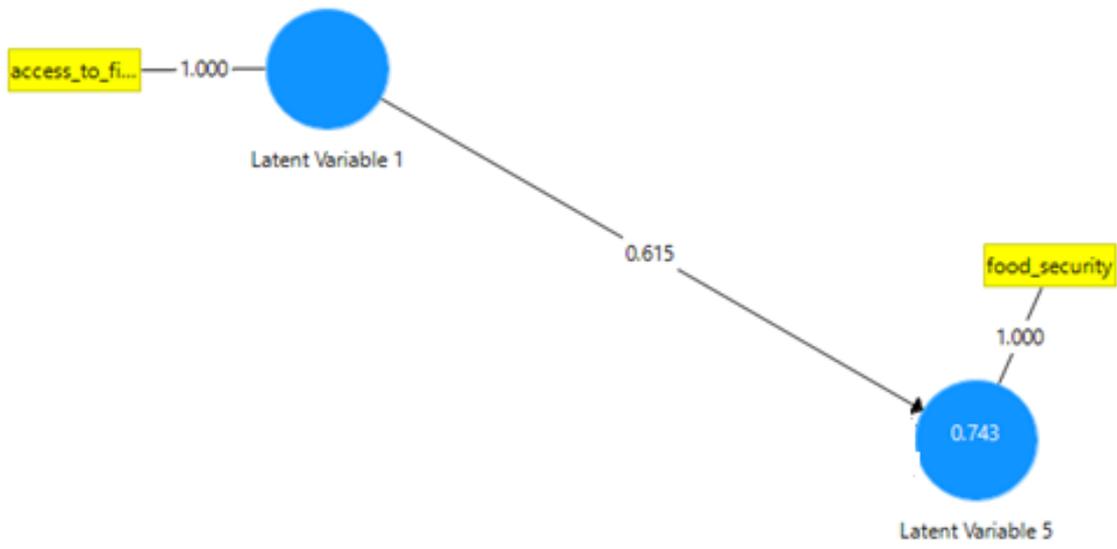

*Fig.2 Measurement Model (Algorithm testing)*

*Source: Author's Field Survey, 2024*

Table 1 provides a detailed explanation of Figure 4.1, which depicts the path coefficients, t-values, and standard error at which the hypotheses were supported or rejected. The t-values in this study were computed using 5000 resampling iterations in repetitive bootstrapping. The selection of 5000 samples is justified by the fact that it assures that each model parameter has an empirical sampling distribution, and the distribution's standard deviation serves as a proxy for the parameter's empirical standard error (Hair et al., 2012).

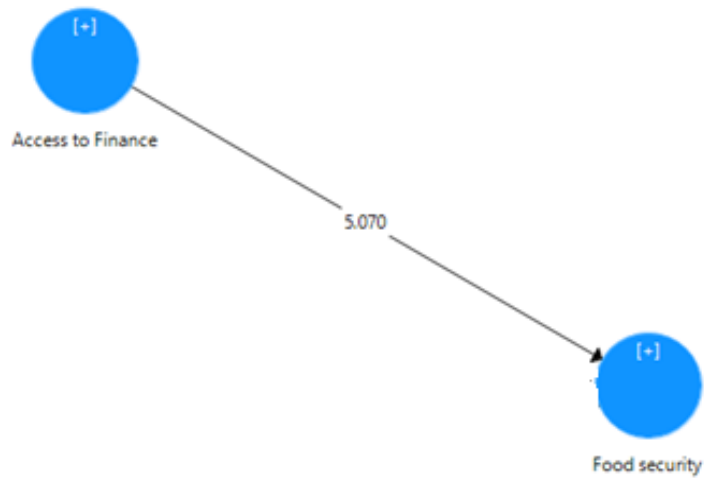

Fig. 3 Structural Model (Bootstrapping @5000) Food Security

*Source: Author's Field Survey, 2024*

Table 1

*Structural model result for the food security*

| Constructs | Original Sample (O) | Sample Mean (M) | Standard Deviation (STDEV) | T Statistics (|O/STDEV|) | P Values |
|---|---|---|---|---|---|
| Access to Finance -> Food security | 0.615 | 0.574 | 0.121 | 5.070 | 0.000 |

*Source: Author's Field Survey, 2024*

In Hypothesis One, the Structural Equation Modelling (SEM) results show that access to finance, as an independent construct, has a positive association with food security. Results (Table 4.1, Figure 4.1) reveal a substantial positive connection between access to finance and food security ($\beta = 0.615$, $t = 5.070$, $p < 0.000$). Improved access to financial resources leads to higher food security levels, with a statistically significant $R^2$ value of 61.5%.

Empirical research backs up this conclusion, as access to money is widely recognised as a significant element impacting food security in a variety of scenarios. For example, Adebayo et al. (2021) discovered that increasing financial access allows smallholder farmers to invest in better agricultural methods, resulting in higher crop yields and greater food availability. Similarly, Asfaw et al. (2020) found that access to financial services not only enables farmers to acquire inputs, but also provides them with the liquidity they need to manage risks connected with agricultural production, hence improving food security.

Furthermore, research from Sulaiman et al. (2022) supports the link between access to finance and food security, arguing that financial inclusion initiatives can significantly reduce food insecurity by empowering households to engage in productive activities and diversify their income sources. The ability to invest in technology and sustainable farming practices is frequently dependent on proper financial assistance, which eventually leads to better food security outcomes.

The favourable impact of access to finance on food security is consistent with the larger theoretical framework of sustainable lifestyles, which holds that financial capital is critical for developing resistance to food insecurity (Scoones, 2021).

This viewpoint emphasises the need of including financial access into food security policies, confirming the idea that increasing financial inclusion can result in significant gains in food security indices. In conclusion, the findings of this study are consistent with current literature, which emphasises the importance of financial access in enhancing food security. The significant relationship ($\beta = 0.615$, $p < 0.000$) suggests that boosting financial access can improve food security outcomes for the target group.

## CONCLUSION

The study revealed that there is a considerable positive association between access to finance and food security, emphasising financial resources as an important aspect in improving agribusiness performance. Improving access to financial services can improve food security outcomes, especially for smallholder farmers with financial restrictions ($R^2 = 61.5\%$). As a result, it is advised that governments develop laws to improve loan disbursement, availability and support targeted financial inclusion programs for the agriculture sector. Furthermore, educating farmers about financial management and accessible goods might help them use these resources more effectively.

Financial institutions should be encouraged to create customised financial solutions, such as low-interest loans and flexible repayment choices, to satisfy the special demands of agribusinesses due to the seasonal and cyclical nature of agriculture. Finally, increasing collaboration between governments, non-governmental organisations, and financial institutions can result in complete support systems that improve financial access and promote sustainable farming practices. Prioritising these techniques allows stakeholders to dramatically improve food security and agribusiness performance in a variety of scenarios.

**РОЗБЛОКУВАННЯ МАЙБУТНЬОГО ПРОДОВОЛЬЧОЇ БЕЗПЕКИ ЧЕРЕЗ ДОСТУП ДО ФІНАНСУВАННЯ ДЛЯ СТАЛОГО РОЗВИТКУ АГРОБІЗНЕСУ**

**Ayobami Paul Abolade**
*Department of Business & Entrepreneurship Kwara State University Malete, Nigeria*

**Ibrahim Olanrewaju Lawal**
*Department of Business & Entrepreneurship Kwara State University Malete, Nigeria*

**Kamoru Lanre Akanbi**
*Department of Business & Entrepreneurship Kwara State University Malete, Nigeria*

**Ahmed Orilonise Salami**
*Africa Business School, UM6P, Rabbat, Morroco*

Доступ до фінансування є життєво важливим для покращення продовольчої безпеки, особливо в країнах, що розвиваються, де сільськогосподарське виробництво має вирішальне значення. Незважаючи на низку фінансових інтервенцій, спрямованих на збільшення сільськогосподарського виробництва, малі фермери продовжують не мати доступу до розумного, своєчасного та достатнього фінансування, що обмежує їхні можливості інвестувати в покращення технологій та виробничих ресурсів, знижуючи продуктивність та постачання продовольства. Це дослідження вивчає взаємозв'язок між доступом до фінансування та продовольчою безпекою серед дрібних фермерів у штаті Огун, використовуючи інституційну теорію як теоретичну основу. У дослідженні використовується кількісний метод з опитуванням 37 200 малих сільськогосподарських фермерів. Розмір вибірки - 380 - був обраний за допомогою імовірнісної вибірки та простих випадкових методів. Дані були проаналізовані за допомогою моделювання структурних рівнянь методом найменших квадратів (PLS-SEM). Отримані результати демонструють сприятливий зв'язок між доступом до фінансування та продовольчою безпекою, причому $R^2$-значення 0,615 вказує на сильний зв'язок. Ці висновки підкреслюють необхідність вдосконалення фінансових установ та впровадження сприятливої політики для забезпечення доступу фермерів до фінансових ресурсів, необхідних їм для досягнення результатів у сфері продовольчої безпеки.

**Ключові слова**: доступ до фінансування, фермер, агробізнес, продовольча безпека.